\documentclass[twocolumn,floats,showpacs,superscriptaddress,pre]{revtex4}
\usepackage{amssymb}
\usepackage{graphicx}
\usepackage{dcolumn}
\usepackage{amsmath}
\usepackage{bm}
\usepackage{epsfig}

\begin{document}

\title{$k$-core (bootstrap) percolation on complex networks:
\\
Critical phenomena and nonlocal effects}
\author{A. V. Goltsev}
\affiliation{Departamento de F{\'\i}sica da Universidade de Aveiro, 3810-193 Aveiro,
Portugal}
\affiliation{A.F. Ioffe Physico-Technical Institute, 194021 St. Petersburg, Russia}
\author{S. N. Dorogovtsev}
\affiliation{Departamento de F{\'\i}sica da Universidade de Aveiro, 3810-193 Aveiro,
Portugal}
\affiliation{A.F. Ioffe Physico-Technical Institute, 194021 St. Petersburg, Russia}
\author{J. F. F. Mendes}
\affiliation{Departamento de F{\'\i}sica da Universidade de Aveiro, 3810-193 Aveiro,
Portugal}

\begin{abstract}
We develop the
theory
of
the
$k$-core (bootstrap) percolation on uncorrelated random
networks with arbitrary degree distributions. We show that the
$k$-core percolation is an unusual, hybrid phase transition with a
jump emergence of the $k$-core as at a first order phase
transition but also with a critical singularity as at a continuous
transition.
We describe the properties of the $k$-core, explain the meaning of
the order parameter for the $k$-core percolation, and reveal the
origin of the specific critical phenomena. We demonstrate that a
so-called ``corona'' of the $k$-core plays a crucial role (corona
is a subset of vertices in the $k$-core which have exactly $k$
neighbors in the $k$-core).
It turns out that the $k$-core percolation
threshold is at the same time the percolation
threshold
of finite
corona clusters.
The
mean separation of vertices in corona clusters plays the role of
the correlation length and diverges at the critical point. We show
that a random removal of even one vertex from the $k$-core may
result in the collapse of a vast region of the $k$-core around the
removed vertex. The mean size of this region diverges at the
critical point. We find an exact mapping of the $k$-core
percolation to a model of cooperative relaxation. This model
undergoes
critical relaxation with a divergent rate at some
critical moment.
\end{abstract}

\maketitle





\section{Introduction} \label{intro}

Random damage may crucially change
the structure and function of a network and may even completely
destroy it. The description of the destruction of the complex
network architectures due to damage is a challenging direction in
the multidisciplinary science of networks. Remarkably, much
attention was attracted to the hierarchical organization of
various real-world networks (the Internet, the WWW, cellular
networks, etc.) and extracting and indexing of their highly
interconnected substructures --- $k$-cores, cliques, and others
\cite{aabv05,k05,wa05,dpv05,dgm05}. The question is: how does
random damage change and destroy these substructures, in
particular, the $k$-cores?

The $k$-{\em core} of a network is its largest subgraph whose
vertices have degree at least $k$. In other words, each of
vertices in the $k$-core has at least $k$ nearest neighbors within
this subgraph. The $k$-core of a graph may be obtained in the
following way (the $k$-core algorithm or ``the pruning
rule''). Remove from the graph all vertices of degree less than
$k$. Some of the rest vertices may remain with less than $k$
edges. Then prune these vertices, and so on until no further
pruning is possible. The result, if it exists, is the $k$-core.
The notion of the $k$-core \cite{b84,s83} is a natural
generalization of the giant connected component in the ordinary
percolation \cite {ajb00,cnsw00,nsw01,cah00,cah02,dms01}. The
$k$-core percolation implies the breakdown of 
the 
giant $k$-core at
a threshold concentration of vertices or edges removed at random.
In physics, the $k$-core percolation (the $k$-core percolation) on
a random Bethe lattice was introduced in Ref.~\cite{clr79} for
describing some magnetic materials. The $k$-core percolation on a
random Bethe lattice was used as a basic model of the quadrupolar
ordering in solid $(o-$H$_{2})_{x}(p-$H$_{2})_{1-x}$ mixtures
\cite{apm87}, the rigidity percolation \cite{mdl97}, the jamming
in sphere packing \cite{slc04}, glassy dynamics \cite{sbt05}, etc.
An exact threshold of the emergence of the $k$-core in some basic
random networks was found in Ref.~\cite{psw96,fr04}. Recently, a
general exact solution of the $k$-core problem for damaged and
undamaged uncorrelated networks with arbitrary degree
distributions was obtained in Ref.~\cite{dgm05}.

These investigations revealed that the $k$-core percolation is
featured by an unusual phase transition which differs strongly
from the ordinary percolation. The latter emerges through a
continuous phase transition occurring at a critical concentration
$p_{c}$ (percolation threshold) of the vertex occupation
probability $p$ \cite {ajb00,cnsw00,nsw01,cah00,cah02,dms01}. (A
vertex is occupied with the probability $p$ and is removed with
the complementary probability $Q=1-p$.) At concentrations
$p>p_{c}$ the giant connected component of a network occupies a
finite fraction $M$ of the total number of vertices $N$ in the
network. At $p\rightarrow p_{c}$, $M$ tends to zero, $M\propto
(p-p_{c})^{\beta }$. The standard mean-field exponent $\beta =1$
takes place in networks with a rapidly decaying degree
distribution.
In networks with a scale-free degree distribution $q^{-\gamma }$,
exponent $\beta$ deviates from the mean-field value at
$2<\gamma <4$ \cite{cah02}. In these networks, $p_c=0$ at $\gamma \leq 3$.

The $k$-core percolation for $k\geqslant 3$ demonstrates another
behavior. When $p\rightarrow p_{c}(k)$, the relative size $ M_{k}$
of the giant $k$-core tends to a finite value $M_{k}(p_{c}(k))$,
and at $ p<p_{c}(k)$ the $k$-core is absent.
Note that the
critical concentration $p_{c}(k)$ depends on $k$. In this
respect, the $k$-core percolation looks like a first-order phase
transition with a jump of $M_{k}$ at the critical point. However,
for a first-order phase transition, one would expect that $M_{k}$
is an analytical function of $p$ at $p>p_{c}(k)$. Contrary to
these expectations, $M_{k}$ shows a singular behavior \cite
{clr79,slc04,hs05,dgm05}: $M_{k}(p)-M_{k}(p_{c}(k))\propto
[p-p_{c}(k)]^{1/2}$. Recently, a similar phase transition was
observed by numerical simulations of the random packing of
frictionless particles (jamming transition) \cite{jamming}. 
In the case of the ordinary percolation, the
critical phenomena are related to the divergence of the mean size
of finite clusters (finite connected components). But what is the
origin of critical phenomena for the $k$-core percolation? First
important steps in resolving this question have been made in
Refs.~\cite{slc04,hs05} where the important role of a so-called
``corona'' of the $k$-core was noted. The {\em corona} is a subset
of vertices in the $k$-core which have exactly $k$ nearest
neighbors in the $k$-core.

In the present paper we develop the qualitative and exact
theories of the $k$-core percolation on complex
networks. Our consideration is based on an exact solution of the
$k$-core percolation on uncorrelated networks with arbitrary
degree distributions. Specifically, we use the configuration model
--- the maximally random graphs with a given degree distribution
\cite{bbk72}. In the large network limit, in any finite
neighborhood of a vertex, these graphs have a tree-like local
structure, i.e., --- without loops, see, e.g.,
Ref.~\cite{dms03,bm05}. Note that in tree-like networks, finite
$k$-cores are absent, and so we discuss only the giant $k$-core.

In Sec.~\ref{damage} we present a qualitative picture and
demonstrate that the critical behavior at the $k$-core percolation
threshold is related to the critical behavior of the corona of the
$k$-core. In Sec.~\ref{basic} we overview an exact solution
describing the $k$-core organization. In Sec.~\ref{edges} we
discuss the statistics of edges in a network with the $k$-core and
the meaning of the order parameter for the $k$-core percolation.
The critical behavior of the order parameter is considered in
Sec.~\ref{point}. Using generating functions, in Sec.~\ref{gf} we
show that the $k$-core percolation threshold is at the
same time the percolation threshold for corona clusters.
At this point the mean size of corona
clusters diverges. The distribution of corona clusters over sizes
is found in Sec.~\ref{distrib}. Specific correlations between
vertices in the $k$-core are discussed in Sec.~\ref{correlation}.
It is demonstrated that the mean intervertex distance in corona
clusters plays the role of the correlation length. In
Sec.~\ref{longrange} we derive exact equations which describe the
evolution of the degree distribution in the $k$-core with
increasing concentration of randomly removed vertices. It is
demonstrated that a removal of even one vertex may result in a
vast damage of the $k$-core. The ``diameter'' of the damaged
region tends to infinity at the $k$-core percolation threshold. In
Sec.~\ref{mapping} we propose an exact mapping of the $k$-core
percolation to a model of cooperative relaxation. This 
model undergoes
critical relaxation with a divergent rate
at some critical moment of the evolution.

\section{Random damaging the $k$-core} \label{damage}

In this section we 
qualitatively describe 
the $k$-core
percolation in an uncorrelated random network for $k\geq 3$. We
assume that a vertex in the original network is occupied with a
probability $p$. Let $M_{k}$ be the probability that a randomly
chosen vertex in a given network belongs to the $k$-core. The
$k$-core in itself is a random network which consists of vertices
with degrees at least $k$. Corona vertices, i.e. vertices with the
degree $k$, are distributed randomly in the $k$-core. They occupy
a finite part of the $k$-core. We shall show in Sec.~\ref{gf} that
at $p>p_{c}(k)$, the corona consists of only finite clusters, so
that its relative size in the $k$-core is sufficiently small. Any
vertex in the $k$-core may have a link with a corona vertex
belonging to a finite corona cluster.


\begin{figure}
\begin{center}
\scalebox{0.34}{\includegraphics[angle=270]{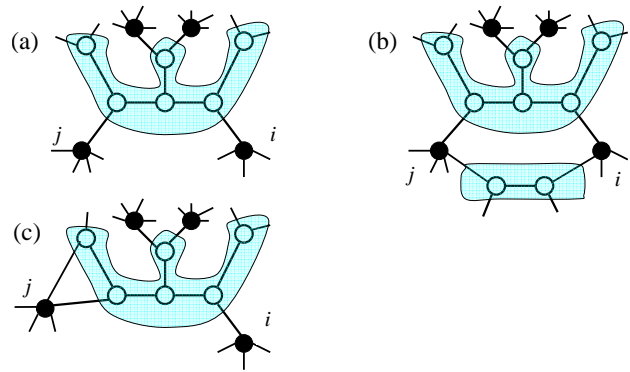}}
\end{center}
\caption{(a) A part of the $k{=}3$-core with a finite corona
cluster which consists of vertices with exactly 3 edges. This
cluster is shown as a grey region. Corona vertices are represented
by open dots. In a tree-like network only one corona cluster may
connect two vertices, for example, vertices $i$ and $j$ on this
figure. Removal of vertex $i$ results in pruning all vertices
which belong to the corona cluster. As a result, vertex $j$ loses
one neighbor (this is the neighboring corona vertex). The degree
of vertex $j$ decreases by 1. (b) In a 
network 
with 
loops, two or more corona clusters may connect together a pair of
vertices in the $k$-core. (c) In networks with nonzero clustering,
a vertex in the $k$-core may be attached to a corona cluster by
two or more edges. In the cases (b) and (c) a removal of the
vertex $i$ results in pruning the corona vertices and, in turn,
the degree of the vertex $j$ is decreased by 2.
}
\label{clusters}
\end{figure}


Let us study the change of the $k$-core size when removing
vertices at random from the original network. Let the occupation
probability $p$ be diminished by a value $\Delta p$. This
corresponds to a random removal of $N\Delta p$ vertices. We
denote the corresponding decrease in the size of the $k$-core by
$N\Delta M_{k}=NM_{k}(p)-NM_{k}(p-\Delta p)$. This change is a
quantity to be found. Firstly, there is a trivial contribution to
$N\Delta M_{k}$ due to the removal of the deleted vertex from the
$k$-core:
\begin{equation}
N\delta M_{k}\equiv N\Delta p\partial M_{k}/\partial p=\Delta pNM_{k}/p
.
\label{pdMk}
\end{equation}
This can be seen from Eqs.~(\ref{Mnk}) and (\ref{k-core}).

Secondly, after removing a vertex together with its edges from the
$k$-core we must prune all other vertices which will occur with
degrees less than $k$. In fact, the removal of a single vertex $i$
from the $k$-core results in the removal of the entire corona
clusters attached to vertex $i$. Note that several corona
clusters may be attached to a vertex with degree $n>k$ in the
$k$-core. The removal of the corona clusters happens due to ``the
domino principle''. Indeed, after removing vertex $i$, its nearest
neighboring corona vertex loses one link with the $k$-core. This
vertex must be pruned from the $k$-core, because now it has only
$k-1$ links with the $k$-core. Due to this removal, each of second
nearest neighbors of vertex $i$ in the corona clusters also loses
one link with the $k$-core and also must be pruned, and so on
until all vertices in the corona clusters will be pruned one by
one. This process is explained in Fig.~\ref{clusters}(a), where a
part of the $k{=}3$-core with a corona cluster is represented. Let
$N_{\text{crn}}$ be the mean total size of all corona clusters 
attached to a vertex in the $k$-core \cite{remark2}. Then the
second contribution to $N\Delta M_{k}$ is $N\delta
M_{k}N_{\text{crn}}$ which is the number of the deleted vertices
in the $k$-core multiplied by $N_{\text{crn}}$.

What happens with other vertices remaining in the $k$-core after
the removal of vertex $i$ together with corona clusters attached
to $i$? If there is no loops, all nearest neighbors of the deleted
vertex $i$ and of the pruned corona vertices remain in the
$k$-core. Their degrees decrease by 1 since each of these vertices
loses one link with the $k$-core: $n\rightarrow n-1\geq k$. On the
other hand, in networks with loops, due to the pruning, a vertex
may lose more than one connection to the $k$-core. Such situations
are represented in Figs.~\ref{clusters}(b) and \ref{clusters}(c).

Thus, in a tree-like network,
\begin{equation}
N\Delta M_{k}=N\delta M_{k}+N\delta M_{k}N_{\text{crn}}
.
\label{deltaMk}
\end{equation}
In a differential form this equation looks as follows:
\begin{equation}
\frac{d\ln M_{k}}{d\ln p}=1+N_{\text{crn}}
.
\label{dM/dp}
\end{equation}
We will show in Sec.~\ref{gf} that corona clusters percolate
exactly at the $k$-core percolation threshold $p_{c}(k)$ and that
$N_{\text{crn}}$ diverges as $[p-p_{c}(k)]^{-1/2}$. Consequently,
according to Eq.~(\ref{dM/dp}), $M_{k}$ demonstrates a critical
singularity: 
\begin{equation}
M_{k}(p)-M_{k}(p_{c}(k))\propto [p-p_{c}(k)]^{1/2} . \label{Mpc}
\end{equation}%
In Sec.~\ref{gf} we will also show that Eq.~(\ref{dM/dp}) is exact
for uncorrelated networks with an arbitrary degree distribution
(the configuration model) in the limit $N\rightarrow \infty $.

\section{Basic equations} \label{basic}  

In this section we 
develop 
an exact formalism for calculating
various $k$-core's characteristics. This approach is based on our
paper \cite{dgm05}.

We consider an uncorrelated network with a given degree
distribution $P(q)$ --- the so-called configuration model. We
assume that a vertex in the network is occupied with the
probability $p$. In this tree-like network, the giant $k$-core
coincides with the infinite $(k{-}1)$-ary subtree. By definition,
the $m$-ary tree is a tree, where all vertices have branching at
least $m$. The introduction of the $(k{-}1)$-ary tree notion
allows one to strictly define the order parameter in the $k$-core
problem for tree-like networks (see below).

\begin{figure}
\epsfxsize=50mm \centerline{\epsffile{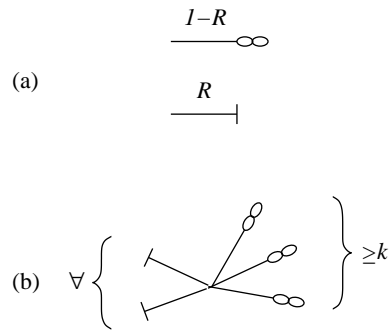}}
\caption{ (a) Schematic representation of the order parameter.
$1-R$ is the probability that a given end of a randomly chosen
edge in an undamaged network is a root of an infinite
$(k{-1})$-ary subtree. $R$ is the probability that a given end of
an edge is not a root of an infinite $(k{-}1)$-ary subtree. (b)
Schematic view of vertex configurations contributing to $M_k$
which is the probability that a vertex is in the $k$-core [see
Eqs.~(\protect\ref{Mnk}) and (\protect\ref{k-core})]. A vertex in
a tree-like network belongs to the $k$-core if at least $k$ its
nearest neighbors are the roots of infinite $(k{-}1)$-ary
subtrees.
The symbol $\forall$ shows that the number of
nearest neighbors which are not roots of infinite $(k{-}1)$-ary
subtrees is arbitrary.
}
\label{f1}
\end{figure}

Let $R$ be the probability that a given end of an edge of a
network is not the root of an infinite $(k{-}1)$-ary subtree.
Then, the probability $M_{k}(n)$ that a vertex chosen at random
has exactly $n\geqslant k$ neighbors in the $k$-core is given by
the equation:
\begin{equation}
M_{k}(n)=p\sum\limits_{q\geqslant
n}^{{}}P(q)C_{n}^{q}R^{q-n}(1-R)^{n}
.
\label{Mnk}
\end{equation}%
Here, $P(q)$ is the probability that a randomly chosen vertex in
the original undamaged network has degree $q$. $C_{n}^{m}R^{q-n}(1-R)^{n}$ is
the probability that a vertex with degree $q$ has $q-n$ neighbors
which are not roots of infinite $(k{-}1)$-ary subtrees
and $n$ neighbors which
are roots of infinite $(k{-}1)$-ary subtrees.
The combinatorial multiplier
$C_{n}^{m}=m!/(m-n)!n!$ gives the number of ways one can choose
$n$ neighbors from $q$ neighbors. A vertex belongs to the $k$-core
if at least $k$ its neighbors are roots of infinite $(k{-}1)$-ary
subtrees. So the probability $M_{k}$ that a vertex belongs to the
$k$-core is equal to 
\begin{equation}
M_{k}=\sum\limits_{n\geqslant k}^{{}}M_{k}(n). \label{k-core}
\end{equation}%
Schematic Fig.~\ref{f1} explains (\ref{Mnk}) and (\ref{k-core}).
Note that for the ordinary percolation we must set $k=1$ in this
equation. The number of vertices in the $k$-core is equal to
$NM_{k}$.


The probability $R$ plays the role of the order parameter in our
problem. Due to using the $(k{-}1)$-ary trees, $R$ is defined in
such a way that it is independent of the second end of the edge
--- whether it belongs or does not belong to the $k$-core.

\begin{figure}
\epsfxsize=82mm \centerline{\epsffile{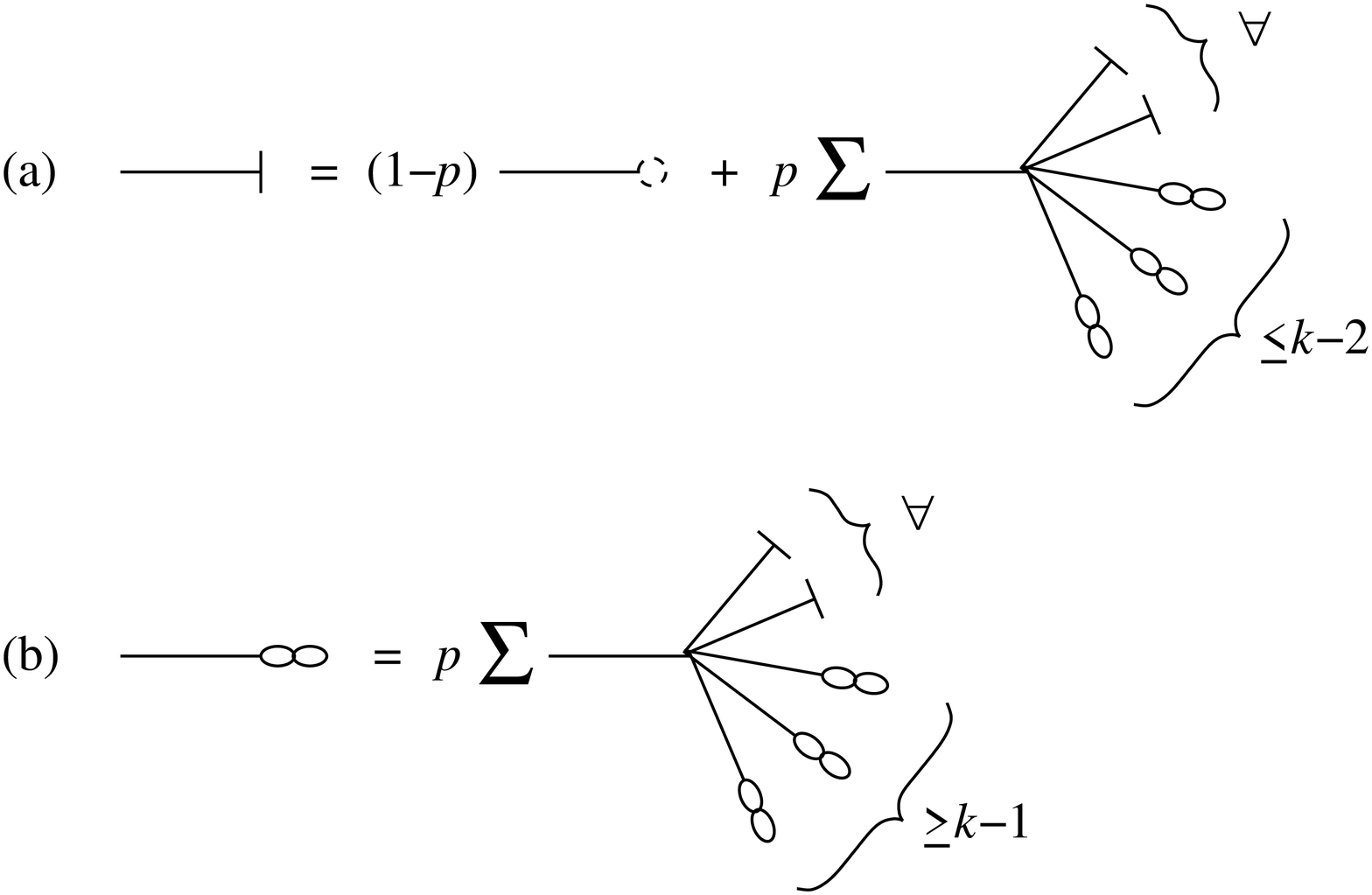}}
\caption{ 
(a) and (b) are graphic representations of
Eqs.~(\protect\ref{R}) and (\protect\ref{Ralt}), respectively. In
(a) the open circle with a dashed boundary represents an
unoccupied vertex. Other notations are explained in the caption to
Fig.~(\ref{f1}).
} 
\label{f2}
\end{figure}

An end of an edge is not a root of an infinite $(k{-}1)$-ary
subtree if at most $k{-}2$ its children branches are roots of
infinite $(k{-}1)$-ary subtrees. This leads to the following exact
equation for $R$ \cite{dgm05}:
\begin{equation}
\!R=1{-}p{+}p\sum_{n=0}^{k-2}\left[ \,\sum_{i=n}^{\infty }\frac{(i{+}1)P(i{+}%
1)}{z_{1}}\,C_{n}^{i}R^{i-n}(1{-}R)^{n}\right] \!\!{.}\!\!\!  \label{R}
\end{equation}%
Let us explain this equation. (i) The first term, $1{-}p\equiv Q$,
is the probability that the end of the edge is unoccupied. (ii)
$C_{n}^{i}R^{i-n}(1-R)^{n}$ is the probability that if a given end
of the edge has $i$ children (i.e., other edges than the starting
edge), then exactly $n$ of them are roots of infinite
$(k{-}1)$-ary subtrees. $(i+1)P(i+1)/z_{1}$ is the probability
that a randomly chosen edge leads to a vertex with branching $i$.
$z_{1}=\sum\nolimits_{q}qP(q)$ is the mean number of the nearest
neighbors of a vertex in the graph. Thus, in the square brackets,
we present the probability that a given end of the edge has
exactly $n$ edges, which are roots of infinite $(k-1)$-ary
subtrees. (iii) Finally, we take into account that $n$ must be at
most $k-2$. In an alternative form, Eq.~(\ref{R}) may be written as
follows,
\begin{equation}
\!1-R=p\sum_{n=k-1}^{\infty}\left[ \,\sum_{i=n}^{\infty }\frac{(i{+}1)P(i{+}%
1)}{z_{1}}\,C_{n}^{i}R^{i-n}(1{-}R)^{n}\right]
\!\!{.}\!\!\!
\label{Ralt}
\end{equation}
This equation shows that a given end of an edge is a root of an
infinite $k-1$-ary tree with the probability $1-R$ if it has at
least $k-1$ children (we sum over $n\geq k-1$) which also are
roots of an infinite $(k{-}1)$-ary subtree. Equations~(\ref{R})
and (\ref{Ralt}) are graphically represented in Fig.~\ref{f2}.


Introducing a function
\begin{equation}
\Phi _{k}(R)\!\!=\sum_{n=0}^{k-2}\sum_{i=n}^{\infty
}\frac{(i{+}1)P(i{+}1)}{z_{1}}\,C_{n}^{i}R^{i-n}(1{-}R)^{n},
\label{F1}
\end{equation}
we rewrite Eq.~(\ref{R}) in a concise form:
\begin{equation}
R=1-p+p\Phi _{k}(R).
\label{R2}
\end{equation}%
If this equation has only the trivial solution $R\!=\!1$, there is
no giant $k$-core. The emergence of a nontrivial solution
corresponds to the birth of the giant $k$-core. The $k$-core is
described by the lowest nontrivial solution $R\!<1$.

The structure of the $k$-core is essentially determined by its
degree distribution $P_{k}(n)$:
\begin{equation}
P_{k}(n)\equiv \frac{M_{k}(n)}{M_{k}}.
\label{Pkq}
\end{equation}
$P_{k}(n)$ is the probability to find a vertex with degree $n$ in
the $k$-core. Note that the $k$-core of an uncorrelated network in
itself is an uncorrelated graph, and so it is completely described
by its degree distribution $P_{k}(n)$ \cite{remark3}. We will
extensively use this circumstance. The corona occupies a fraction
$P_{k}(k)$ of the $k$-core. Therefore, the total number of
vertices in the corona is equal to $NM_{k}P_{k}(k)$. The mean
degree of vertices in the $k$-core is
\begin{equation}
z_{1k}=\sum_{n\geq k}P_{k}(n)n.
\label{z1k}
\end{equation}
Comparing Eqs.~(\ref{z1k}) and (\ref{Ralt}), we get an important
relationship between $z_{1k}$, $M_{k}$ and $1-R$:
\begin{equation}
z_{1k}M_{k}=z_{1}(1-R)^{2}.
\label{z1k2}
\end{equation}
Below, in Sec.~\ref{edges}, we will discuss its meaning.

In the general case the analytical solution of Eq.~(\ref{R}) is
unknown. But it can be obtained numerically. By using this
solution, one can calculate all basic characteristics of the
$k$-core structure of an arbitrary uncorrelated network
\cite{dgm05}.


\begin{figure}
\begin{center}
\scalebox{0.35}{\includegraphics[angle=270]{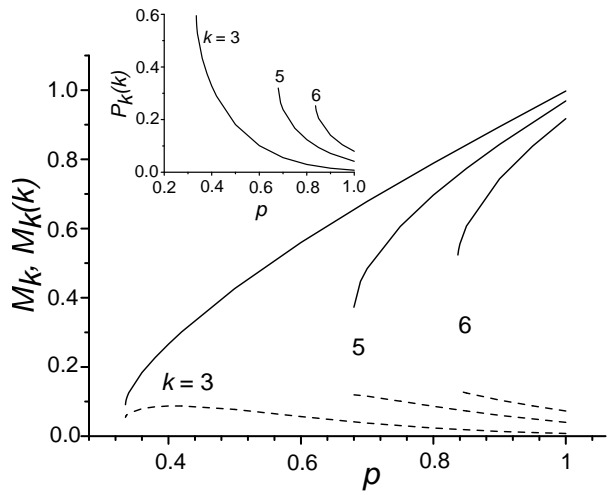}}
\end{center}
\caption{ Dependence of the sizes of the $k$-core and the corona,
$M_{k}$ and $M_{k}(k)$, respectively, on the occupation
probability $p$ in the Erd\H os-R\'{e}nyi graphs with the mean
degree $z_1=10$. Solid lines show $M_{k}$ versus $p$, and dashed
lines show $M_{k}(k)$ versus $p$ at $k=3,5,$ and 6. Notice that both
$M_{k}$ and $M_{k}(k)$ have a square root singularity at the
$k$-core percolation thresholds, but in the curves $M_{k=5,6}(k)$
this singular addition is practically unnoticeable. Notice the
non-monotonous dependence $M_{k}(k)$. The inset shows the fraction
$P_{k}(k)$ of the corona vertices in the $k$-core versus $p$. }
\label{crn-size}
\end{figure}


Some results of the numerical solution of Eq.~(\ref{R}) for the
Erd\H os-R\'{e}nyi graph with $z_1=10$ are represented in
Fig.~\ref{crn-size}. More results may be found in
Ref.~\cite{dgm05}. This figure displays the dependences of the
sizes of the $k$-core, the corona and the fraction of the corona
vertices in the $k$-core on the occupation probability $p$ for
several values of $k$. One can see that far from the critical
point $p_{c}(k)$ the size of the corona is small in comparison to
the size of the $k$-core. However, close to $p_{c}(k)$ the corona
occupies a noticeable fraction of the $k$-core.

Let us consider a network with a scale-free degree distribution,
$P(q)\propto q^{-\gamma}$. The case $2<\gamma \leqslant 3$ is
realized in most important real-world networks. With $\gamma $ in
this range, the mean number of the nearest neighbors of a vertex
in a network, $z_{2}$, diverges if $N\rightarrow \infty $. Solving
analytically Eq.~(\ref{R2}), we find that the size of the
$k$-core decreases with increasing $k$ as follows~\cite{dgm05}:
\begin{equation}
M_{k}=p^{2/(3-\gamma )}(q_{0}/k)^{(\gamma -1)/(3-\gamma )},
\label{k-size}
\end{equation}
where $q_{0}$ is the minimum degree in the initial (undamaged)
network. Vertices which belong to the $k$-core, but do not belong
to the $ k{+}1$-core, form the $k$-shell of the size
$S_{k}=M_{k}-M_{k+1}$. Using Eq.~(\ref{k-size}), at $k\gg 1$ we
find: 
\begin{equation}
S_{k}\propto (q_{0}/k)^{2/(3-\gamma )}.
\label{shell-size}
\end{equation}
The asymptotic behavior given by Eqs.~(\ref{k-size}) and
(\ref{shell-size}) agrees well with an empirical analysis of the
$k$-core architecture of the Internet on the Autonomous Systems
level \cite{aabv05,k05}.

\section{Statistics of edges and the order parameter} \label{edges}

\begin{figure}
\epsfxsize=34mm
\centerline{\epsffile{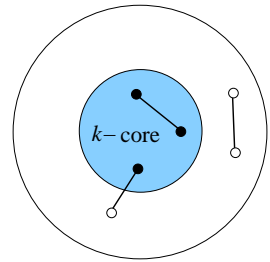}}
\caption{
Schematic
representation of the three types of edges in a network (large
circle) with the $k$-core (grey central region): (i) edges between
vertices in the $k$-core (links between two black dots), (ii)
edges between vertices which do not belong to the $k$-core (links
between two open dots), and (iii) edges between vertices in the
$k$-core and vertices which do not belong to the $k$-core (links
between black and open dots).
}
\label{nedges}
\end{figure}

\begin{figure}[b]
\epsfxsize=82mm
\centerline{\epsffile{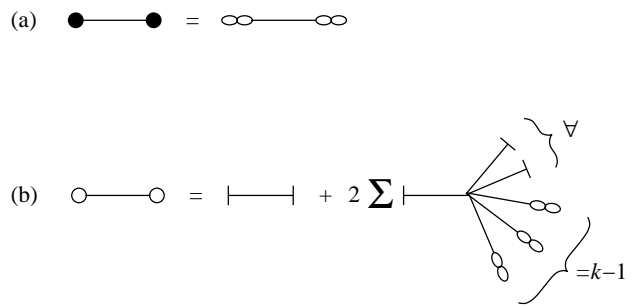}}
\caption{
Schematic representations of the probabilities that an
edge connects together vertices of the $k$-core or that it
connects vertices outside the $k$-core, figures (a) and (b),
respectively.
}
\label{f6}
\end{figure}

Let us consider edges in an uncorrelated network with the
$k$-core.
We start with the case $p=1$.
There are three types of edges: (i) edges which connect two
vertices in the $k$-core, (ii) edges connecting two vertices which
do not belong to the $k$-core, and (iii) edges connecting together
a 
vertex 
in 
the $k$-core and the other one which do not belong
to the $k$-core. These types of connections in a network are
schematically shown in Fig.~\ref{nedges}. Let $L_{k}$, $L_{0}$ and
$L_{0k}$ be the total numbers of edges of these three types in the
network, respectively. The sum of these numbers gives the total
number $L=Nz_{1}/2$ of edges in the initial network: 
\begin{equation}
L_{0}+L_{k}+L_{0k}=L
.
\label{Lt}
\end{equation}
The  ratios $L_{k}/L$, $L_{0}/L$ and $L_{0k}/L$ are probabilities
that a randomly chosen edge in the initial network is of the type
(i), (ii) or (iii), respectively. Because $L_{k}=Nz_{1k}M_{k}/2$,
we can rewrite Eq.~(\ref{z1k2}) in a form: 
\begin{equation}
\frac {L_{k}}{L}=(1-R)^{2}.
\label{z1k3}
\end{equation}
This equation has a simple explanation. 
It 
shows that the
probability to find an edge which connects two vertices in the
$k$-core is equal to the probability that both its ends are
roots of the $(k{-}1)$-ary tree, that is, $(1-R)^2$
[see Fig.~\ref{f6}(a)]. On the other hand, Eq.~(\ref{z1k3})
explains the meaning of the order parameter $R$ via the
relationship with the measurable parameters: $1-R=\sqrt{L_{k}/L}$.

One should note that Eq.~(\ref{z1k3}) is also
valid at $p<1$
since
it follows from the exact equation~(\ref{z1k2}). In this case,
$L_{k}$ must be replaced
by the number $L_{k}(p)$ of edges in a damaged $k$-core while $L$
remains
the total number of edges in the initial network.


Let us find the probability $L_{0}/L$ that an edge chosen at
random in the network connects two vertices which do not belong to
the $k$-core. We stress that $L_{0}/L$ is not equal to $R^2$ as
one could naively expect, but
is
larger,
see Fig.~\ref{f6}(b).
Indeed, in addition to configurations where both the ends of an
edge are not the roots of infinite $(k{-}1)$-ary trees --- the
$R^2$ contribution --- one must take into account extra
configurations. In these additional configurations, one end of the
edge is not the root of an infinite $(k{-}1)$-ary tree, but the
second end has exactly $k-1$ childrens which are roots of infinite
$(k{-}1)$-ary trees.
This second vertex does not belong to the
$k$-core as it should be.
Thus we have
\begin{equation}
L_{0}/L = R^2+2R\sum\limits_{q\geqslant
k}^{{}}\frac{qP(q)}{z_{1}}C_{k-1}^{q-1}R^{q-k}(1-R)^{k-1}.
\label{L01}
\end{equation}
Comparing the sum in Eq.~(\ref{L01})
and the probability $M_{k}(k)$ given by Eq.~(\ref{Mnk}) at $n=k$,
we get
\begin{equation}
L_{0}/L=R^2+2R\frac{kM_{k}(k)}{z_{1}(1-R)}.
\label{L0}
\end{equation}
Equations (\ref{Lt}), (\ref{z1k3}) and (\ref{L0}) establish
nontrivial relationships between independently measurable network
parameters: $L$, $L_{k}$, $L_{0}$, $L_{0k}$, and $M_{k}(k)$. These
relations may be used as a criterion of the validity of the tree
ansatz for various networks.

Let us now touch upon the case $p<1$.
After random removal vertices from an uncorrelated network, we again get an uncorrelated network.
Therefore, at $p<1$, we may still use the same formulas~(\ref{z1k3}), (\ref{L01}), and (\ref{L0}) but with substituted characteristics of the damaged network --- the number of edges, the mean degree, etc.~\cite{remark5}.


\section{$k$-core percolation threshold} \label{point}

When decreasing the occupation probability $p$, the $k$-core
decreases in size and disappears at a critical concentration
$p_{c}(k)$. According to Ref.~\cite{dgm05}, the critical
concentration $p_{c}(k)$ is determined by the following equation:
\begin{equation}
p_{c}(k)\Phi _{k}^{\prime }(R_{c})=1.
\label{cp1}
\end{equation}
Here, $R_{c}$ is a critical value of the order parameter $R$ at
the birth point of the $k$-core. At $p<p_{c}(k)$, Eq. (\ref{R2})
has only the trivial solution $R=1$, and the giant $k$-core does
not exist. The derivative $\Phi _{k}^{\prime }(R)\equiv d\Phi
_{k}(R)/dR$ is determined by the following equation:
\begin{eqnarray}
p\Phi _{k}^{\prime }(R) &= &p \sum\limits_{q\geqslant k}^{{}}\frac{qP(q)}{%
z_{1}}C_{k-2}^{q-1}(q+1-k)R^{q-k}(1-R)^{k-2}
\notag
\\[5pt]
&=&k(k-1)P_{k}(k)/z_{1k}
.
\label{F2}
\end{eqnarray}
Using this equation, the condition (\ref{cp1}) for the $k$-core
percolation threshold may be rewritten in the form:
\begin{equation}
k(k-1)P_{k}(k)/z_{1k}=1.
\label{cp2}
\end{equation}

Let us consider the behavior of $R$ near the phase transition in
an uncorrelated complex network with a finite mean number $z_{2}$
of the second neighbors of a vertex,
$z_{2}=\sum\nolimits_{q}q(q-1)P(q)$. At $p$ near $p_{c}(k)$, i.e.,
at $0<p-p_{c}(k)\ll 1$, Eq.~(\ref{R}) has a nontrivial solution:
\begin{equation}
R_{c}-R\propto \lbrack p-p_{c}(k)]^{1/2}.
\label{expR}
\end{equation}
This demonstrates that $R$ has a jump at $p=$ $p_{c}(k)$ [from
$R=R_{c}$ to $R=1$] as at an ordinary first-order phase transition
and a singular behavior as at a continuous phase transition
\cite{clr79}. The derivative $dR/dp$ diverges, 
\begin{equation}
\frac{dR}{dp}=-\frac{1-R}{p[1-p\Phi _{k}^{\prime }(R)]}\propto
-[p-p_{c}(k)]^{-1/2}
,
\label{dR}
\end{equation}
since at $p\rightarrow p_{c}(k)+0$ we have 
\begin{equation}
1-p\Phi _{k}^{\prime }(R)\propto [p-p_{c}(k)]^{1/2}
.
\label{expFi}
\end{equation}
This singularity suggests intriguing critical phenomena near the
threshold of the $k$-core percolation.

In contrast, in networks with infinite $z_{2}$, instead of the
hybrid phase transition, the $k$-core percolation becomes an
infinite order phase transition \cite{dgm05}, similarly to the
ordinary percolation in this situation \cite {cnsw00}.\ In this
case the entire $k$-core organization of a network is extremely
robust against random damage.

\section{Generating functions for corona clusters} \label{gf}

Using the approach of Refs. \cite{cnsw00,nsw01}, we introduce the
generating function $H_{1}(x)$ of the probability that an end of a
randomly chosen edge in the $k$-core belongs to a finite corona
cluster of a given size: 
\begin{equation}
H_{1k}(x)=1-\frac{kP_{k}(k)}{z_{1k}}+x\frac{kP_{k}(k)}{z_{1k}}%
[H_{1k}(x)]^{k-1}
.
\label{H1}
\end{equation}
Here $kP_{k}(k)/z_{1k}$ is the probability that an end of an edge
chosen at random in the $k$-core belongs to the corona. In turn,
$1-kP_{k}(k)/z_{1k}$ is the complementary probability that the end
of the edge does not belong to the corona. We have $H_{1k}(1)=1$.

We introduce the generating function $H_{0k}(x)$ for the size of a
corona cluster attached to a vertex in the $k$-core:
\begin{equation}
H_{0k}(x)=\sum\limits_{q}P_{k}(q)[H_{1k}(x)]^{q}
.
\label{H0}
\end{equation}
Using this function, one can calculate the mean total size
$N_{\text{crn}}$ of the corona clusters attached to a vertex
randomly chosen in the $k$-core: 
\begin{equation}
N_{\text{crn}}=\left.\frac{dH_{0k}(x)}{dx}\right\vert _{x=1}
.
\label{Ncrn1}
\end{equation}
Differentiating 
Eqs.~(\ref{H1}) and (\ref{H0}) over $x$ gives
\begin{equation}
N_{\text{crn}}=\frac{kP_{k}(k)}{1-k(k-1)P_{k}(k)z_{1k}^{-1}}
.
\label{Ncrn2}
\end{equation}

Inserting Eqs.~(\ref{F2}) and (\ref{expFi}) into
Eq.~(\ref{Ncrn2}), we find that $N_{\text{crn}}$ diverges at the
critical point,
\begin{equation}
N_{\text{crn}}\propto \lbrack p-p_{c}(k)]^{-1/2}
.
\label{Ncrn3}
\end{equation}
This means that at $p=p_{c}(k)$ the corona is in its ``percolation
transition threshold''. Note however that the $k$-core and its
corona are absent at $p<p_{c}(k)$, so that there is no giant
connected corona above this threshold, in contrast to the ordinary
percolation. Equation (\ref{cp2}) resembles the condition
$p z_{2}/z_{1}=1$ of the emergence of the giant connected component
in the ordinary percolation.

The exact derivation of Eq.~(\ref{dM/dp}) is based on the
following steps. We differentiate Eq.~(\ref{k-core}) for $M_{k}$
over $p$:
\begin{equation}
\frac{dM_{k}}{dp}=\frac{M_{k}}{p}-kP_{k}(k)\frac{M_{k}}{(1-R)}\frac{dR}{dp}
.
\label{dM/dp2}
\end{equation}
Inserting Eqs.~(\ref{F2}), (\ref{dR}) and (\ref{Ncrn2}) into
Eq.~(\ref{dM/dp2}), we get Eq.~(\ref{dM/dp}).

\section{Size distribution of corona clusters} \label{distrib}

Let $\mathfrak{N}_{\text{crn}}(s)$ be the number of corona
clusters of 
size $s$ in the $k$-core. Because the total
number of vertices in the corona clusters is equal to
$NM_{k}P_{k}(k)$, we obtain the following condition:
\begin{equation}
\sum\limits_{s=1}^{s_{\max}}s\mathfrak{N}_{\text{crn}}(s)=NM_{k}P_{k}(k)
,
\label{Ns1}
\end{equation}
where $s_{\max}$ is the size of the largest corona cluster. We
introduce a function
\begin{equation}
\Pi _{k}(s)\equiv \frac{s\mathfrak{N}_{\text{crn}}(s)}{NM_{k}P_{k}(k)
},
\label{Ps0}
\end{equation}
which is the probability that a randomly chosen corona vertex
belongs to a corona cluster of 
size $s$. The function $\Pi
_{k}(s)$ is related to the generating function
$G_{\text{crn}}(x)$:
\begin{equation}
\Pi _{k}(s)=\frac{1}{s!}\left.
\frac{d^{s}G_{\text{crn}}(x)}{dx^{s}}\right\vert _{x=0}
,
\label{PS1}
\end{equation}
where $G_{\text{crn}}(x)=x[H_{1k}(x)]^k$. There is a simple
relationship between $N_{\text{crn}}$ and the mean size
$s_{\text{crn}}$ of a corona cluster to which a randomly chosen
corona vertex belongs: 
\begin{equation}
s_{\text{crn}}\equiv\sum\limits_{s=1}^{s_{\max}}s\Pi
_{k}(s)=1+\frac{k}{z_{1k}}N_{\text{crn}}
. 
\label{crnM}
\end{equation}
At $s\gg 1$ the probability $\Pi _{k}(s)$ has the usual asymptotic
form \cite{nsw01}: 
\begin{equation}
\Pi _{k}(s)\approx Cs^{-\alpha }e^{-s/s^{\ast }}
.  
\label{PS2}
\end{equation}
Here $C$ is a constant. Exponent $\alpha$ and the parameter
$s^{\ast }=1/\ln \left\vert x^{\ast }\right\vert $ are determined
by the
type and the position $x^{\ast }$ of the singularity of the
function $H_{0k}(x)$, nearest to $x=1$. Solving Eq.~(\ref{H1}) and
inserting the obtained solution into Eq.~(\ref{H0}), we find that
at $p\rightarrow p_{c}(k)+0$ the generating functions $H_{1k}(x)$
and $G_{\text{crn}}(x)$ have a square-root singularity: 
\begin{equation}
G_{\text{crn}}(x)\propto H_{1k}(x)\propto (1-x)^{1/2} 
.
\label{sH01}
\end{equation}
In the case $k=3$, Eq.~(\ref{H1}) is solved exactly,
\begin{equation}
H_{1k=3}(x)=\frac{1-(1-x/x^{\ast })^{1/2}}{2ax}
,
\label{sH3}
\end{equation}
where $a=3P_{3}(3)/z_{1,k=3}$ and $x^{\ast }=1/[4a(1-a)]$. At the
critical point, Eq. (\ref{cp2}), we have $2a=1$ and, therefore,
$x^{\ast }=1$. At $p$ near $p_{c}(k)$, the parameter $s^{\ast }$
diverges, 
\begin{equation}
s^{\ast }\approx 1/(1-2a)^{2} \propto 1/[p-p_{c}(k)] .
\label{star}
\end{equation}
The singularity  
(\ref{sH01}) results in the standard
mean-field exponent $\alpha =3/2$. At the critical point
$p=p_{c}(k)$, the distribution function is
\begin{equation}
\Pi _{k}(s)\propto s^{-3/2}
.
\label{e39}
\end{equation}
It gives $\mathfrak{N}_{\text{crn}}(s) \propto \Pi _{k}(s)/s \sim s^{-5/2}$.
In scale-free networks with a degree distribution $P(q) \sim
q^{-\gamma}$, this is valid for any $\gamma>3$. In contrast, in
the ordinary percolation on scale-free uncorrelated networks,
exponent $\alpha $ differs from the standard mean-field value
$3/2$ if $2<\gamma <4$ \cite{cah02,cha03}.

Let us estimate the
size $s_{\max }$ of the largest corona
cluster. We use the condition that there is only one corona
cluster with the size $s\geq s_{\max }$:
\begin{equation}
\int\limits_{s_{\max }}^{\infty }\mathfrak{N}_{\text{crn}}(s)ds=NM_{k}P_{k}(k)%
\int\limits_{s_{\max }}^{\infty }\Pi _{k}(s)s^{-1}ds=1
.
\label{smax}
\end{equation}
Using asymptotics (\ref{PS2}), at $p>p_{c}(k)$ we obtain
\begin{equation}
s_{\max }\propto \ln N
.
\label{e41}
\end{equation}
At the critical point $p=p_{c}(k)$, using the distribution
function $\Pi _{k}(s)\propto s^{-3/2}$, we get
\begin{eqnarray}
s_{\max} & \propto & N^{2/3}
,
\label{e42}
\\[5pt]
s_{\text{crn}} & \propto & \sqrt{s_{\max}} \propto N^{1/3}
.
\label{critCRN}
\end{eqnarray}
Equation (\ref{e42}) coincides with a result for the maximum size
of a connected component at the birth point of the giant connected
component in classical random graphs \cite{bollobasbook} and in
uncorrelated networks where three first moments of a degree
distribution converge \cite{cha03}. However, relation~(\ref{e42})
essentially differs from that for the maximum size of a connected
component if the third moment of the degree distribution diverges
in the infinite network limit.

\begin{figure}
\epsfxsize=48mm
\centerline{\epsffile{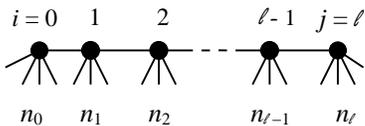}} \caption{Diagrammatic
representation of the mean number
$\mathcal{P_{\ell}}(n_{0},n_{1},\ldots,n_{\ell})$ of ways to reach
a vertex which is at 
distance $\ell$ from a vertex $i=0$ in
the $k$-core. The path goes through vertices
$m=1,2,\ldots,\ell{-}1$ with degrees $n_{m}$ in the $k$-core.}
\label{chain}
\end{figure}


\section{The correlation length} \label{correlation}

In this section we consider correlations between vertices in the
$k$-core with $k\geq 3$. Let us chose at random a vertex $i$ in
the $k$-core. We aim to find the mean number $\mathcal{P_{\ell}}$
of vertices in the $k$-core which may be reached from $i$
following 
a path of 
a 
length $\ell $.
In the configuration model the giant $k$-core is a unique and
simply connected subgraph. Therefore, all $\ell -1$ vertices on a
path connecting $i$ and $j$ must also belong to the $k$-core.
$\mathcal{P_{\ell}}$ is given by the following 
relation: 
\begin{eqnarray}
&& \mathcal{P_{\ell}}(n_{0},n_{1},\ldots, n_{\ell }) \nonumber
\\[5pt]
&&\!\!\!\!\!\!\!\!\!\!\!\!\!\!\!\! =P_{k}(n_{0})n_{0} \prod\limits_{m=1}^{\ell
-1}\Bigg[\frac{P_{k}(n_{m})n_{m}(n_{m}-1)}{z_{1k}}\Bigg] \frac{P_{k}(n_{\ell
})n_{\ell }}{z_{1k}}
.
\label{Ppath}
\end{eqnarray}
A diagrammatic representation of $\mathcal{P_{\ell}}$ is shown in
Fig.~\ref{chain}. Here $n_{m}$ is the degree of vertex $m$, where
$m=0,1...$ $\ell $, on a path connecting $i$ and $j$ in the
$k$-core. For 
the sake of 
convenience we set $i\equiv 0$ and $j\equiv \ell $.
Here $P_{k}(n_{0})$ is the probability that $i$ has the degree
$n_{0}$ in the $k$-core. The multiplier $n_{0}$ gives the number
of ways to reach vertex $1$ from $i{=}0$ following along any of
its $n_{0}$ edges. In the brackets, $P_{k}(n_{m})n_{m}/z_{1k}$ is
the probability that an edge outgoing from a vertex $m-1$ leads to
a vertex $m$ of degree $n_{m}$. The multiplier $n_{m}-1$ gives the
number of ways to leave this vertex.
Finally, $P_{k}(n_{\ell })n_{\ell }/z_{1k}$ is the probability
that the final edge on the path leads to the destination vertex
$\ell $ of degree $n_{\ell }$.

Now it is easy to find the number $N_{\text{crn}}(\ell )$ of
corona vertices which are at a distance $\ell$ from a randomly
chosen vertex in the $k$-core and belong to corona clusters
attached to this vertex. We set $n_{1}=n_{2}=...=n_{\ell }=k$ and
sum over degree $n_{0}$ of the starting vertex $0$ in
Eq.~(\ref{Ppath}). Using Eq.~(\ref{z1k}) gives 
\begin{eqnarray}
N_{\text{crn}}(\ell )&&=\sum\limits_{n_{0}\geq k}
\mathcal{P_{\ell}}(n_{0},n_{1}=k,...n_{\ell}=k) \notag
\\[5pt]
&&=kP_{k}(k)e^{-(\ell -1)/\lambda }
.
\label{Ncrn4}
\end{eqnarray}
Here we have introduced the correlation length
\begin{equation}
\lambda \equiv -\,\frac{1}{\ln [p\Phi _{k}^{\prime }(R)]} =
-\,\frac{1}{\ln[k(k-1)P_{k}(k)/z_{1k}]}
.
\label{lambda}
\end{equation}%
In accordance with Eq.~(\ref{expFi}), at $p\rightarrow p_{c}(k)+0$
the parameter $\lambda$ diverges,
\begin{equation}
\lambda \propto [p-p_{c}(k)]^{-1/2} . \label{lambda2}
\end{equation}
Summing $N_{\text{crn}}(\ell)$ over $\ell$, we reproduce
Eq.~(\ref{Ncrn2}) for $N_{\text{crn}}$:
\begin{equation}
N_{\text{crn}}=\sum_{\ell =1}^{\infty }N_{\text{crn}}(\ell
)=\frac{kP_{k}(k)}{1-\exp [-1/\lambda ]}
.
\label{Ncrn5}
\end{equation}
Let us determine the mean intervertex distance $r_{\text{crn}}(k)$
between vertices in corona clusters. We use the quantity
$\mathcal{P_{\ell}}(n_{0},n_{1},\ldots,n_{\ell})$ and set
$n_{0}=n_{1}=\ldots=n_{\ell}=k$. We find 
\begin{eqnarray}
r_{\text{crn}}(k) &\equiv &\frac{\sum_{\ell =1}^{\infty } \ell
\mathcal{P_{\ell}}(k,k,\ldots, k)}{\sum_{\ell =1}^{\infty
}\mathcal{P_{\ell}}(k,k,\ldots, k)} \notag
\\[5pt]
&=&\frac{1}{1-\exp [-1/\lambda ]} . \label{Radius}
\end{eqnarray}
At $p$ close to $p_{c}(k)$, the correlation length $\lambda \gg 1$
and therefore $r_{\text{crn}}(k)\approx \lambda $.

\section{Nonlocal effects in the $k$-core percolation} \label{longrange}

In Sec.~\ref{damage} we have shown that a removal of a vertex from
the $k$-core leads to pruning corona clusters attached to the
vertex. In this section we will demonstrate that a removal of even
one vertex from the $k$-core, $k\geq 3$, influences degrees of
vertices in a vast region of the $k$-core around this vertex.
Moreover, the size of this damaged region diverges at the critical
point.

Let $N\Delta p$ vertices be removed at random from the initial
network.
As a result, the total number of vertices with degree $n$ in the
$k$-core is changed, 
\begin{equation}
N\Delta M_{k}(n)=NM_{k}(p,n)-NM_{k}(p-\Delta p,n). \label{DMkn1}
\end{equation}
Let us find $N\Delta M_{k}(n)$. With probability $M_{k}(n)$, a
removed vertex may have degree $n$ in the $k$-core. Therefore,
there is a trivial contribution to $N\Delta M_{k}(n)$: 
\begin{equation}
N\delta M_{k}(n)=N\Delta p\partial M_{k}(n)/\partial p=N\Delta pM_{k}(n)/p
.
\label{Mn1}
\end{equation}
Removal of a vertex $i$ in the $k$-core may influence on the
degree of a vertex $j$ which is at a distance $\ell $ from $i$. If
$j$ is a nearest neighbor of $i$, then the degree $n$ of vertex $j$ will
be decreased by 1. If $\ell >1$, then the probability of this
effect is determined by the probability that $j$ and $i$ are
connected by a chain of corona vertices. If vertex $i$ is removed,
then all vertices of a corona cluster attached to $i$ also must be
pruned from the $k$-core due to the domino principle. As a result,
vertex $j$ loses one neighbor in the $k$-core. Let $V(\ell ,n)$ be
the mean number of vertices of degree $n$ which are connected to a
randomly chosen vertex $i$ in the $k$-core by a chain of corona
vertices of length $\ell $. Removal at random $N\delta M_{k}$
vertices results in a decrease of $M_{k}(n)$ by a quantity 
\begin{equation}
N\delta M^{(2)}(n)=N\delta M_{k}\sum _{\ell =1}^{\infty }V(\ell
,n)
.
\label{Mn2}
\end{equation}
At the same time vertices with degree $n+1$ may also lose one edge
with the $k$-core. After the pruning, they have $n$ edges within
the $k$-core. This effect increases $M_{k}(n)$ by a quantity
\begin{equation}
N\delta M^{(3)}(n)=-N\delta M_{k}\sum _{\ell =1}^{\infty }V(\ell
,n+1)
.
\label{Mn3}
\end{equation}

Note that only in networks with loops, vertices in the $k$-core may change their degree by 2 during the pruning. 
Thus, in a tree-like network there
are only three contributions to $N\Delta M_{k}(n)$:
\begin{eqnarray}
N\Delta M_{k}(n) &=&N\delta M_{k}(n)+N\delta M_{k}\sum _{\ell
=1}^{\infty
}V(\ell ,n)
\notag
\\[5pt]
&&-N\delta M_{k}\sum _{\ell =1}^{\infty }V(\ell ,n+1)
.
\label{deltaMkn}
\end{eqnarray}
$V(\ell ,n)$ is given by the probability $\mathcal{P_{\ell }}$,
Eq.~(\ref{Ppath}):
\begin{eqnarray}
V(\ell ,n)&=&\sum\limits_{n_{0}\geq
k}\mathcal{P_{\ell}}(n_{0},n_{1}=k,\ldots,n_{\ell -1}=k, n_{\ell
}=n) \nonumber
\\[5pt]
&=&nP_{k}(n)e^{-(\ell -1)/\lambda } . \label{Vln}
\end{eqnarray}
Inserting this result into Eq.~(\ref{deltaMkn}), in the limit
$\Delta p\rightarrow 0$, we get the main result of the present
section: 
\begin{equation}
\frac{dM_{k}(n)}{d\ln p}=M_{k}(n)+rnM_{k}(n)-r(n+1)M_{k}(n+1)
,
\label{dM/dp3}
\end{equation}
where
\begin{equation}
r=\frac{1}{1-\exp [-1/\lambda ]}=\left[
1-\frac{k(k-1)M_{k}(k)}{\sum\nolimits_{n=k}^{q_{\text{cut}}}nM_{k}(n)}\right]
^{-1}.
\label{rs}
\end{equation}
The parameter $r$ determines the mean size of a region in the
$k$-core which is damaged by a removal of one vertex chosen at
random. One should stress that $r$ depends on the entire degree
distribution in the $k$-core: $r=r\{M_k(n)\}$. At $p$ close to
$p_{c}(k)$, we have $r\propto \lambda$. Therefore, at
$p\rightarrow p_{c}(k)$, this size diverges. Interestingly, the
parameter $r$ is equal to the mean intervertex distance
$r_{\text{crn}}$ in corona clusters given by 
expression~(\ref{Radius}).

In Eq.~(\ref{dM/dp3}), the index $n$ can take the values
$n=k,k+1,\ldots,q_{\text{cut}}$. The cutoff $q_{\text{cut}}$ of
the network's degree distribution $P(q)$ depends on details of a
specific network and its size $N$.

Although we derived Eq.~(\ref{dM/dp3}) by using heuristic
arguments, this equation is exact for uncorrelated random graphs
in the limit $N\rightarrow \infty$. Equation~(\ref{dM/dp3}) may be
strictly derived by differentiating Eq.~(\ref{Mnk}) over $p$ and
using Eq.~(\ref{dR}).

The set of Eqs.~(\ref{dM/dp3}) with $n$ from $k$ to
$q_{\text{cut}}$ is a complete set of nonlinear equations which
determine $M_{k}(n)$ as a function of $p$. The nonlinearity is due
to the functional dependence of $r$ on $M_{k}(n)$.

Summing over $n$ from $k$ to $q_{\text{cut}}$ on the left and right hand
sides of Eq.~(\ref{dM/dp3}), we obtain Eq.~(\ref{dM/dp}). If we
know $M_{k}(n)$ for an initial network, i.e., at $p=1$, then we
can use Eq.~(\ref{dM/dp3}) and find the evolution of $M_{k}(n,p)$
with decreasing $p$. Inserting Eqs.~(\ref{Pkq}) and (\ref{z1k})
into (\ref{z1k2}), we can determine the order parameter $R$ as a
function of $p$,
\begin{equation}
R=1-\Bigg[z_{1}^{-1}\sum\limits_{n=k}^{q_{\text{cut}}}nM_{k}(n,p)\Bigg]^{1/2}
.
\label{R3}
\end{equation}
Alternatively, we could find the order parameter $R(p)$, solving
Eq.~(\ref{R}), and afterwards obtain $M_{k}(n)$ from
Eq.~(\ref{Mnk}).


\begin{figure}
\epsfxsize=80mm
\centerline{\epsffile{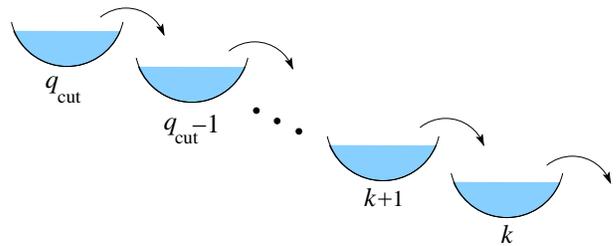}}
\caption{
Schematic picture
of a relaxation process described by Eq.~(\protect\ref{dM/dt}). An
initial distribution $M_{k}(n,t=0)$ over states with
$n=q_{\text{cut}},q_{\text{cut}}-1,\ldots,k$ relaxes into the
final state $\{M_{k}(n)=0 \}$ due to transitions of vertices from
a state with degree $n+1$ to a state with degree $n$. Here
$q_{\text{cut}}$ is the maximum degree.
}
\label{relaxation}
\end{figure}


\section{Mapping to a cooperative relaxation model} \label{mapping}

Let us consider the $k$-core percolation as an evolutionary
process. At time $t=0 $ we have an initial uncorrelated network
with the $k$-core. During a time interval $\Delta t$ we remove at
random a fraction $\Delta p/p=\Delta t$ of occupied vertices from
the network. This means that the occupation probability $p$
decreases in time as $p=e^{-t}$. With this substitution,
Eq.~(\ref{dM/dp3}) takes a form
\begin{equation}
\frac{dM_{k}(n,t)}{dt}=-M_{k}(n,t)-rnM_{k}(n,t)+r(n+1)M_{k}(n+1,t)
.
\label{dM/dt}
\end{equation}
This rate equation describes the relaxation of an initial
distribution $\{M_{k}(n,t=0)\}$ to the final state with the
destroyed $k$-core, i.e., $\{M_{k}(n)=0\}$, due to the chain of
transitions of vertices from states of degree $n+1$ to states of
degree lower by one: $n+1 \to n$, see Fig.~\ref{relaxation}. Note
that we consider only relaxation in states with $n \geq k$,
assuming that vertices of degree less than $k$ are pruned
instantly \cite{remark4}. The parameter $r$ plays the role of the
characteristic scale of the relaxation rate. This relaxation is a
cooperative process due to the functional dependence of $r$ on
$M_{k}(n,t)$, see Eq.~(\ref{rs}). At time
$t_{c}(k)=\ln[1/p_{c}(k)]$ this model undergoes a dynamical phase
transition. Using Eq.~(\ref{Mpc}), we find that the total number
$M_{k}(t)$ of vertices in the $k$-core has a singular time
dependence near $t_{c}(k)$: 
\begin{equation}
M_{k}(t)-M_{k}(t_{c}(k))\propto \lbrack p(t)-p_{c}(k)]^{\nu
}\propto [t_{c}(k)-t]^{\nu } 
. 
\label{Mtc}
\end{equation}
The critical exponent $\nu =1/2$ is valid for $k\geqslant 3$.
Inser\-ting Eq.~(\ref{lambda2}) into Eq.~(\ref{rs}), we find that
the relaxation rate diverges at the critical time $t_{c}(k)$,
\begin{equation}
r\propto \lbrack p(t)-p_{c}(k)]^{-1/2}\propto [t_{c}(k)-t]^{-1/2}
.
\end{equation}
Note that in accordance with the results obtained in
Sec.~\ref{damage}, the characteristic scale $r$ of the relaxation
rate also determines the mean size of the region in the $k$-core
cropped out due to the deletion of a vertex. In its turn, $r$ is
approximately equal to the correlation length $\lambda$, i.e., the
larger is the correlation length the larger is the relaxation
rate. This is in contrast to the usual critical slowing down of
the order parameter relaxation for continuous phase transitions.
In the latter case, the larger is the correlation length the
smaller is the relaxation rate, $r\approx \lambda ^{-z}$, where
$z$ is a dynamical critical exponent.

\section{Conclusions}\label{conclusion}

In this paper we have explained the nature of the $k$-core
percolation transition in uncorrelated networks.
To
obtain our results, we used heuristic arguments and developed
an exact theory. Let us list the main features
of the quite unusual $k$-core percolation transition: (i) a jump
emergence of the $k$-core, (ii) the critical singularity in the
phase with the $k$-core, (iii) the absence of any critical effects
--- strong correlations, divergent ``susceptibilities'', etc. ---
on the ``normal phase'' side. We had to reveal the meaning of the
order parameter in this problem, to explain the nature of the
jump, to find the origin of the singularity, to indicate a
``physical'' quantity, which diverges at the critical point, to
indicate long-range correlations, specific for this transition.

We have shown that the order parameter in this problem is simply expressed in terms of the relative number of edges in the $k$-core, see relation~(\ref{Lt}). The tree ansatz has allowed us to find the $k$-core order parameter and other $k$-core characteristics of various uncorrelated networks.

We have found that the unique properties of the $k$-core percolation transition are essentially determined by the corona subset of the $k$-core, that is, by vertices with exactly $k$ connections to the $k$-core. These are the ``weakest'' vertices in the $k$-core.
The critical correlations in the $k$-core are due to the correlations in the system of the corona clusters.

In the ``$k$-core phase'', the corona clusters are finite, but
their sizes and long-range correlations grow as the network
approaches the $k$-core percolation threshold. The mean size of a
corona cluster to which a randomly chosen corona vertex belong
diverges at the $k$-core percolation threshold. This quantity
plays the role of a critical susceptibility in this problem. So,
the $k$-core percolation threshold coincides with the percolation
threshold for corona clusters, and the $k$-core phase is the
``normal'' phase for the corona. The dramatic difference from the
ordinary percolation is that the corona disappears on the other
side of the threshold, and so critical fluctuations in the phase
without the $k$-core are absent.

For understanding the nature of this transition, we have studied the process of the destruction of the $k$-core due to the random deletion of vertices.
The deletion of a vertex in the $k$-core results in the clipping out the entire adjacent corona clusters from the $k$-core due to the domino principle. This effect is enormously increased when corona clusters become large --- near the $k$-core percolation threshold. In the threshold, the removal of a tiny fraction of vertices results in the complete collapse of the corona and the $k$-core.
In this respect, the $k$-core percolation problem can be mapped to a model of cooperative relaxation, which undergoes critical relaxation with a divergent rate at some critical moment.

To conclude, let us indicate a possible application --- a social
network model,
where social links connect individuals. Each of vertices (individuals) in our model may
occur in one of a few states --- distinct beliefs, opinions, religions,
ideologies, diseases, etc. We assume that each vertex takes a
specific state if at least $k$ its neighbors are in this state. Is
it possible that in this social net a giant, say, religious group
will emerge? The answer is yes if the network has the giant
$k$-core. A giant community of individuals being in the same state
forms the $k$-core of this
network.
We believe that our results are applicable to a variety of complex cooperative systems of this kind.

\begin{acknowledgments}

This work was partially supported by projects POCTI: FAT/46241,
MAT/46176, FIS/61665, and BIA-BCM/62662, and DYSONET. Authors
thank J.G.~Oliveira for help in numerical calculations.

\end{acknowledgments}




\begin{thebibliography}{10}
\bibitem{remark2}
Note that vertex $i$ is not included in the number $N_{\text{crn}}$ even if this vertex belongs to the corona.
\bibitem{remark3}
This important statement can be proved by the direct calculation of the conventional probability $P_k(n'|n)$ that if a vertex in the $k$-core is of degree $n$, then its nearest neighbor in the $k$-core subgraph is of degree $n'$. This probability turns out to be independent of $n$.
\bibitem{remark5}
Note that in this formulation, the value of the order parameter [let us denote it by $\widetilde{R}(p)$] turns out to be different from $R(p)$ defined in Sec.~\protect\ref{basic}. The reason is that in the definition of $R(p)$ for a damaged network, the edges of the original undamaged net were used [see, e.g., Fig.~\protect\ref{f1}(a)]. So in the damaged network, these edges may be absent. These two order parameter values are simply related to each other: $1-R(p)=\sqrt{L(p)/L}[1-\widetilde{R}(p)]$, where $L(p)$ is the total number of edges in the damaged net. This trivial renormalization directly follows from relation~(\protect\ref{Lt}).
\bibitem{remark4} The $k$-core algorithm, which we use to extract
the $k$-core from a complex network, may be written in the form of
a nonlinear rate equation similar to Eq.~(\protect\ref{dM/dt}), see Refs.~\cite{w95,hw06}.
\end{thebibliography}
\end{document}